\documentclass[pdflatex,sn-basic,Numbered]{sn-jnl}

\usepackage{graphicx}%
\usepackage{multirow}%
\usepackage{amsmath,amssymb,amsfonts}%
\usepackage{amsthm}%
\usepackage{mathrsfs}%
\usepackage[title]{appendix}%
\usepackage{xcolor}%
\usepackage{textcomp}%
\usepackage{manyfoot}%
\usepackage{booktabs}%
\usepackage{algorithm}%
\usepackage{algorithmicx}%
\usepackage{algpseudocode}%
\usepackage{listings}%
\usepackage{lmodern}
\usepackage{anyfontsize}
\usepackage{hyperref}

\begin{document}

\title[Article Title]{Perception of an AI Teammate in an Embodied Control Task Affects Team Performance, Reflected in Human Teammates’ Behaviors and Physiological Responses}



\author*[1]{\fnm{Yinuo} \sur{Qin}}\email{yinuo.qin@columbia.edu}

\author[1,2]{\fnm{Richard} \spfx{T} \sur{Lee}}\email{rtl2118@columbia.edu}

\author*[1,2,3]{\fnm{Paul} \sur{Sajda}}\email{psajda@columbia.edu}

\affil[1]{\orgdiv{Department of Biomedical Engineering}, \orgname{Columbia University}, \orgaddress{\city{New York}, \state{NY}, \country{USA}}}

\affil[2]{\orgdiv{Department of Electrical Engineering}, \orgname{Columbia University}, \orgaddress{\city{New York}, \state{NY}, \country{USA}}}

\affil[3]{\orgdiv{Department of Radiology}, \orgname{Columbia University}, \orgaddress{\city{New York}, \state{NY}, \country{USA}}}

\abstract{The integration of artificial intelligence (AI) into human teams is widely expected to enhance performance and collaboration. However, our study reveals a striking and counterintuitive result: human-AI teams performed worse than human-only teams, especially when task difficulty increased. Using a virtual reality-based sensorimotor task, we observed that the inclusion of an active human-like AI teammate disrupted team dynamics, leading to elevated arousal, reduced engagement, and diminished communication intensity among human participants. These effects persisted even as the human teammates' perception of the AI teammate improved over time. These findings challenge prevailing assumptions about the benefits of AI in team settings and highlight the critical need for human-centered AI design to mitigate adverse physiological and behavioral impacts, ensuring more effective human-AI collaboration.}

\maketitle
\newpage
\section{Introduction}\label{intro}

The rapid progress in artificial intelligence (AI) has revolutionized various facets of human society, enabling a wide range of applications that were once confined to science fiction. From autonomous systems in transportation to decision-support tools in healthcare and finance, AI has seamlessly integrated into our daily lives, reshaping how we work, interact, and make decisions \cite{silver2018general, berner2019dota, topol2019high, mckinney2020international, ouyang2020video}. This growing ubiquity highlights the urgent need to study how human behaviors and interactions are influenced by the integration of AI into collaborative and decision-making processes.

A long-standing hypothesis within the AI community is that effective AI partners in human teams should possess two critical attributes: human-like behavior and super-human intelligence \cite{weber2011building, korteling2021human, meta2022human, duenez2023social}. AI agents with these qualities are believed to foster trust and improve team performance, as they align closely with human expectations of collaboration. Past research has supported this notion by demonstrating that AI systems equipped with advanced capabilities -- such as theory of mind and natural communication -- can significantly enhance human trust and performance in human-AI partnerships \cite{wang2021towards, williams2022supporting, westby2023collective}. This has led to the widely accepted perception that AI agents serve as a ``second brain'' to augment human cognitive abilities and strengthen human-AI collaborations.

Empirical evidence has shown that human-AI teams often outperform teams composed solely of humans or AI agents, particularly in tasks that require complementary strengths \cite{bansal2019beyond, passalacqua2024practice}. However, these studies predominantly focus on dyadic interactions, where a single human collaborates with a single AI agent acting as an assistant to be queried or consulted \cite{lee2022coauthor, wu2022ai, sharma2023human, glickman2024human, vaccaro2024combinations}. As AI technology advances, a foreseeable future involves AI agents becoming active teammates within larger human teams, playing more participatory and collaborative roles. Despite this, significant gaps remain in understanding how the inclusion of AI as an active team member impacts team dynamics, individual behaviors, and physiological responses \cite{zhang2021ideal, reverberi2022experimental, lematta2022remote, schelble2024towards}.

Moreover, while prior research has explored human-AI interactions in domains like text-based communication and decision-making \cite{lee2022coauthor, wu2022ai, sharma2023human, vaccaro2024combinations}, there is a conspicuous lack of studies examining embodied collaborative tasks. Embodied AI, which interacts physically and cognitively with humans in shared environments, represents a frontier application of AI in critical areas such as autonomous driving, manufacturing, and healthcare. These contexts demand nuanced collaboration and high-stakes decision-making, raising questions about how AI agents affect human teams operating in such environments.

In this study, we address these gaps by conducting the first investigation of embodied collaborative control tasks to examine how human-like AI teammates influence team performance, cognitive dynamics, behavior, and physiological responses. Our experimental design leverages the Apollo Distributed Control Task (ADCT), a complex sensorimotor task performed in a virtual reality (VR) environment. Participants collaborated in triads, comprising either all-human teams or teams with a Wizard of Oz (WOz) AI agent -- an expert human-controlled system presented as an advanced AI partner \cite{maulsby1993prototyping}. We collected and analyzed multi-modal data, including team performance metrics, motor actions, verbal communication, physiological signals (i.e., pupil size, blink rate, and electroencephalography (EEG)), and self-reported evaluations of team dynamics.

Surprisingly, our findings challenge prevailing assumptions about the benefits of human-AI collaboration. Contrary to expectations, teams involving a WOz AI agent exhibited significantly lower performance compared to all-human teams. This decline was accompanied by pronounced changes in human behavior and physiology. Participants working with the AI agent experienced heightened arousal, as indicated by increased pupil dilation and blink rates, alongside reduced engagement and communication intensity. These physiological and behavioral changes disrupted team dynamics, making effective collaboration more challenging. Furthermore, while participants’ perceptions of the AI agent improved over time in terms of leadership, helpfulness, and trust, team performance remained suboptimal. Our results underscore the complex interplay between human subjective perception, behavior, and performance.

These results highlight a critical, yet often overlooked, consideration in the design of AI agents for collaborative tasks: the impact of AI presence on human cognition and physiology. Our study highlights the importance of minimizing negative effects on human teammates, both to enhance performance and to promote sustainable and effective collaboration. We argue that the next generation of AI agents must not only optimize their individual capabilities but also account for the cognitive and physiological states of their human partners. By prioritizing these considerations, AI systems can better support human teams in achieving their collective goals.

\begin{figure}[btp]
    \centering
    \includegraphics[width=1\textwidth]{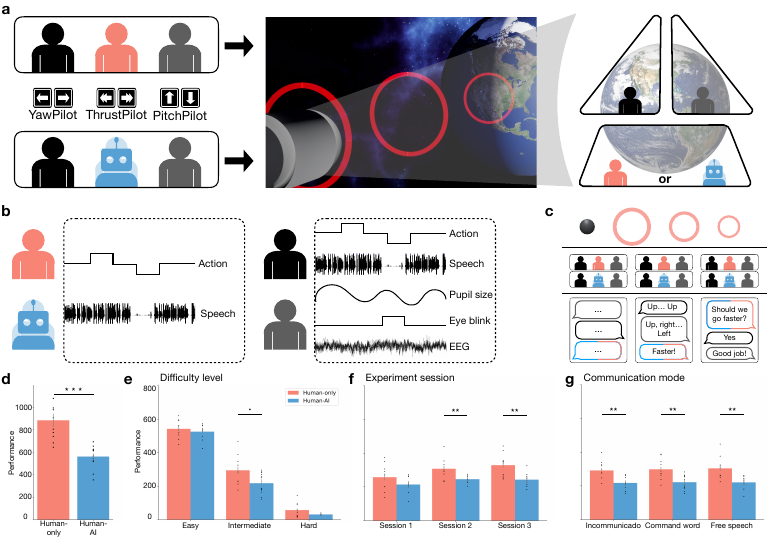}

\caption{Apollo Distributed Control Task (ADCT) overview and performance analysis. 
\textbf{a}, Experiment setup for human-only and human-AI teams. The left panel shows the different team configurations and participant roles. In human-AI teams, the AI agent always acts as the ThrustPilot. The middle panel illustrates the virtual environment where participants collaborate to control a spacecraft, navigating through red rings and returning to Earth. The right panel depicts the relative positioning and partial views for each role. \textbf{b}, Multi-modal data collection for each role. Behavioral data were recorded for the ThrustPilot, while both physiological and behavioral data were recorded for the YawPilot and PitchPilot in both human-only and human-AI teams. \textbf{c}, Three task conditions in this experiment. From top to bottom, our experiment includes three difficulty levels (Easy, Intermediate, Hard), three experimental sessions (Session 1, 2, 3), and three communication protocols (Incommunicado, Command Word, Free Speech) \textbf{d-g}, Team performance under different task conditions, measured by the number of rings passed. Bars represent performance as mean ± s.e.m. Individual dots represent the number of rings passed by each team (human-only: n=10 teams; human-AI: n=12 teams). \textbf{d}, Team performance across all task conditions. \textbf{e}, Team performance within each task difficulty level. The color key for the team is used for f and g. \textbf{f}, Team performance within each experimental session. \textbf{g}, Team performance within each communication protocol. $\cdot~P< 0.1, *~P < 0.05, **~P < 0.01, ***~P < 0.001$ by One-way analysis of variance (ANOVA) test with Bonferroni correction.}
    \vspace{-5pt}
    \label{fig:experiment_fig1}
\end{figure}

\section{Results}\label{results}
We investigated how team dynamics, individual behavior, and physiology differ when team members collaborate with another human participant or a Wizard of Oz (WOz) AI. We designed a two-phase virtual reality-based sensorimotor task called the Apollo Distributed Control Task (ADCT) to achieve this. Our approach provides an immersive environment that facilitates comprehensive physiological data collection while minimizing external distractions \cite{musick2021happens, schoonderwoerd2022design, gu2024data, runzheimer2024exploring}. 

In ADCT, two participants each controlled a different degree of movement of a spacecraft, navigating with a partial view of the environment to safely return to Earth. In the first phase, two human participants collaborated with a third human participant without prior knowledge of the task. In the second phase, two new participants collaborated with a WOz AI named Alice (Fig.~\ref{fig:experiment_fig1} a). Participants were informed that Alice was an AI agent trained using state-of-the-art machine learning models, capable of understanding and communicating with them. However, Alice was controlled by an expert experimenter who was familiar with the task and trained to be consistent throughout trials. By comparing team dynamics, behavior, and physiology between these two phases, we analyzed the differences between all-human teams and teams that included an AI collaborator. This setup allowed us to explore human-AI interaction's effects on team performance, individual actions, and physiological responses.

In each phase of the experiment, we simultaneously collected multi-modal physiological and behavioral data from all participants. The data collected included task performance, speech recordings, pupil size, eye openness, electroencephalography (EEG), and subjective ratings of team members' leadership and helpfulness (Fig.~\ref{fig:experiment_fig1} b). Comparing this data revealed significant changes in team dynamics, individual behavior, and individual physiology when a WOz AI was added to the team. These findings demonstrate the substantial impact of AI collaboration on multiple aspects of team interaction and performance.

To investigate team dynamics, individual behavior, and physiology under different conditions, we varied three key task variables in ADCT: task difficulty level, experimental session, and communication protocol (Fig.~\ref{fig:experiment_fig1} c). Task difficulty was categorized into three levels—easy, intermediate, and hard—based on the size of the ring the spacecraft had to navigate through. The larger the ring radius, the easier the task. The experiment consisted of three sessions, where the same teams participated in three iterations. Three distinct communication protocols were used: 1) incommunicado, where participants were unable to communicate; 2) command word communication, where participants could only use predefined command words; and 3) free speech, where participants could communicate freely. Except for incommunicado trials, participants could communicate at any time, and multiple participants could speak simultaneously. These variables allowed us to systematically examine how task demands and communication protocols influenced the performance of human and human-AI teams.

\begin{figure}[hbtp]
    \centering
    \includegraphics[width=1\textwidth]{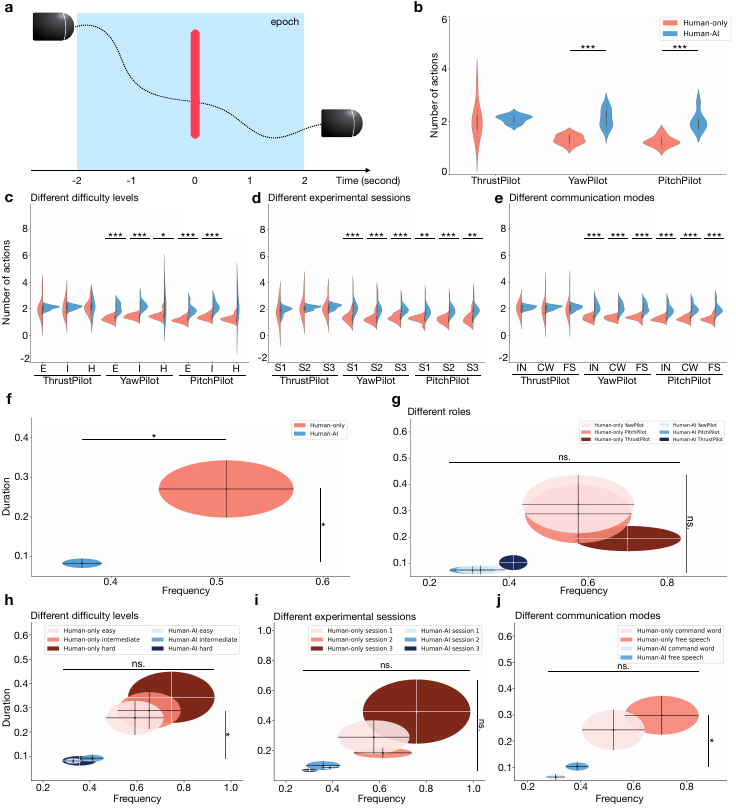}
    \caption{Behavioral data comparison between human-only and human-AI teams. \textbf{a}, Illustration of an analyzed epoch centered around the moment a ring is passed. Each epoch includes two seconds before and after the ring. \textbf{b-e}, Number of remote controller actions in an epoch for each role under different task conditions (human-only: n=10 teams; human-AI: n=12 teams). \textbf{b}, Number of remote controller actions across all conditions. The color key for the team is used for c-e \textbf{c}, the Number of remote controller actions under different task difficulty levels. \textbf{d}, Actions under different experimental sessions. \textbf{e}, Actions under different communication protocols. \textbf{f-j}, Comparison of communication frequency and duration under different task conditions. \textbf{f}, Speech frequency and duration across all roles and task conditions. \textbf{g}, Speech frequency and duration for each role. \textbf{h}, Speech frequency and duration under different task difficulty levels. \textbf{i}, Speech frequency and duration under different experimental sessions. \textbf{j}, Speech frequency and duration under different communication protocols. One-way ANOVA with Bonferroni correction: ns. not significant, $*~P < 0.05, **~P < 0.01, ***~P < 0.001.$}
    \label{fig:behavior_fig2}
\end{figure}

\subsection{Human-Only Teams Outperformed Human-AI Teams in Task Performance}
First, we compare the overall performance of human-only teams and human-AI teams. Team performance was measured by the total number of rings successfully navigated, with possible scores ranging from 0 to 2025. As shown in Fig.~\ref{fig:experiment_fig1} d, all-human teams outperformed human-AI teams, with statistical analysis using one-way analysis of variance (ANOVA) confirming this difference to be significant ($F(1, 20) = 23.278, P = 0.0001$). Even though a human expert controlled the WOz AI agent, the presence of an AI team member led to a decrease in overall team performance.

Next, we explored how team performance varied across different task conditions. For both human-only and human-AI teams, performance dropped significantly as task difficulty increased (Fig.~\ref{fig:experiment_fig1} e, Repeated Measures ANOVA: human-only, $F(2,18) = 563.974, P < 0.0001$; human-AI, $F(2,22) = 734.367, P < 0.0001$). Notably, in easy tasks, human teams did not significantly outperform human-AI teams ($F(1,20) = 0.619; P = 0.4406$). However, as task difficulty increased to intermediate or hard levels, human teams exhibited a much higher performance advantage (intermediate, $F(1,20) = 5.721, P = 0.0267$; hard, $F(1,20) = 4.395, P = 0.0490$). These results suggest that the presence of an AI team member has a more negative effect on performance as tasks become more challenging.

Next, we examined how team performance varied across different task conditions, with a focus on the performance gap between human-only and human-AI teams as task difficulty increased. For both team types, performance dropped significantly as task difficulty increased (Fig.~\ref{fig:experiment_fig1}e, Repeated Measures ANOVA: human-only, $F(2,18) = 563.974, P < 0.0001$; human-AI, $F(2,22) = 734.367, P < 0.0001$). However, the effect of task difficulty on performance differed between the two team types. In easy tasks, human teams did not significantly outperform human-AI teams ($F(1,20) = 0.619; P = 0.4406$). As task difficulty increased to intermediate or hard levels, the performance gap increased, with human teams demonstrating a much larger advantage (intermediate, $F(1,20) = 5.721, P = 0.0267$; hard, $F(1,20) = 4.395, P = 0.0490$). This indicates that the presence of an AI team member has a disproportionately negative effect on performance in more challenging tasks. Hard tasks appear particularly detrimental to human-AI teams due to their high coordination demands and increased cognitive load, which may amplify the difficulties of integrating an AI agent into team workflows. Specifically, the AI’s limitations in adapting to nuanced team communication and human behavioral cues under high stress and complexity likely contribute to this performance gap. The behavioral and physiological analysis that follows further elucidates these findings.

We also analyzed how experimental sessions and communication protocols impacted team performance in human-only and human-AI teams. Unlike task difficulty, experimental sessions and communication protocols had a smaller influence on team performance (Fig.~\ref{fig:experiment_fig1} f and g). Increasing the number of experimental sessions or altering communication protocols did not significantly affect performance in either type of team (Repeated Measures ANOVA: experimental session: human-only, $F(2,18) = 2.793, P = 0.0790$; human-AI, $F(2,22) = 3.056, P = 0.0606$. Communication protocol: human-only, $F(2,18) = 0.100, P = 0.9055$; human-AI, $F(2,22) = 0.036, P = 0.9651$). However, human teams consistently outperformed human-AI teams in the last two sessions and across all communication protocols (session 1, $F(1, 20)=2.723, P = 0.1145$; session 2, $F(1, 20)=11.263, P = 0.0031$; session 3, $F(1, 20)=14.349, P = 0.0012$; incommunicado, $F(1, 20)=13.257, P = 0.0016$; command word, $F(1, 20)=12.061, P = 0.0024$; free speech, $F(1, 20)=13.916, P = 0.0013$). These results suggest that, while communication and repeated collaboration play a role, human teams benefit more from experience compared to human-AI teams.

\subsection{Increased Participant Control Actions When Collaborating with AI Contribute to Reduced Team Performance}
We analyzed individual behaviors by comparing the frequency of remote controller actions in all human teams versus human-AI teams. Data collected in four-second intervals around ring-passing events (Fig.~\ref{fig:behavior_fig2} a) revealed that participants working with the AI agent executed more remote controller actions than those in all-human teams (YawPilot: $F(1, 20) = 44.330, P < 0.0001$; PitchPilot: $F(1, 20) = 36.416, P < 0.0001$), despite the agent’s action frequency being similar to human counterparts ($F(1, 20) = 0.113, P = 0.7408$). This suggests that participants felt a heightened need to assert control or compensate for perceived deficiencies when collaborating with an AI.

Further analysis across varying task difficulties, sessions, and communication protocols consistently showed that participants in human-AI teams performed significantly more control actions than those in all-human teams (Fig.~\ref{fig:behavior_fig2}c-e). Notably, while the agent’s actions remained comparable to those of human participants across conditions, human team members adjusted their behavior differently: they reduced their actions when paired with another human but increased them when paired with the agent (ThrustPilot: easy, $F(1, 20) = 0.233, P = 0.6346$; intermediate, $F(1, 20) = 0.226, P = 0.6394$; hard, $F(1, 20) = 0.006, P = 0.9378$; session 1, $F(1, 20) = 0.256, P = 0.6181$; session 2, $F(1, 20) = 0.007, P = 0.9335$; session 3, $F(1, 20) = 0.419, P = 0.5247$; incommunicado, $F(1, 20) = 0.327, P = 0.5740$; command words, $F(1, 20) = 0.144, P = 0.7083$; free speech, $F(1, 20) = 0.019, P = 0.8923$. YawPilot: easy, $F(1, 20) = 42.428, P < 0.0001$; intermediate $F(1, 20) = 49.834, P = 0.5247$; hard, $F(1, 20) = 4.931, P = 0.0387$, session 1, $F(1, 20) = 19.097, P = 0.0003$; session 2, $F(1, 20) = 18.935, P = 0.0003$; session 3, $F(1, 20) = 25.525, P = 0.0001$; incommunicado, $F(1, 20) = 33.221, P < 0.0001$; command words, $F(1, 20) = 43.152, P < 0.0001$; free speech, $F(1, 20) = 41.376, P < 0.0001$. PitchPilot: easy, $F(1, 20) = 37.153, P < 0.0001$; intermediate, $F(1, 20) = 38.534, P < 0.0001$; hard, $F(1, 20) = 6.087, P = 0.0233$; session 1, $F(1, 20) = 16.022, P = 0.0007$; session 2, $F(1, 20) = 26.479, P < 0.0001$; session 3, $F(1, 20) = 16.362, P = 0.0006$; incommunicado, $F(1, 20) = 29.548, P < 0.0001$; command words, $F(1, 20) = 42.039,P < 0.0001$; free speech, $F(1, 20) = 31.899, P < 0.0001$). This overcompensation likely reflects an attempt to compensate for perceived shortcomings in the AI teammate’s actions, which disrupted team coordination and contributed to reduced overall performance.

The increased frequency of control actions in human-AI teams appears to be closely tied to the decreased team performance observed under these conditions. Heightened arousal, as indicated by physiological measures, has been linked to overcompensation behaviors that negatively impact performance in human-AI collaborations \cite{tyson1998physiological, faller2019regulation, oz2023role}. These findings suggest that the presence of an AI teammate induces significant behavioral changes, such as increased control actions, which may amplify coordination inefficiencies and hinder overall team effectiveness. By integrating behavioral and physiological data, we gain deeper insight into the challenges of human-AI collaboration and the specific factors contributing to performance gaps.

\subsection{Reduced Communication in Human-AI Teams as a Potential Contributor to Lower Performance}
Effective communication is crucial for successful collaboration. Our analysis compared communication frequency and duration between human-only and human-AI teams under various conditions. Findings indicate that human-AI teams communicated less frequently and for shorter durations than human-only teams (Fig.~\ref{fig:behavior_fig2}f; frequency: $F(1, 20)=4.445, P = 0.0478$; duration: $F(1, 20)=7.059, P = 0.0151$), suggesting that the introduction of AI alters team communication dynamics.

Examining communication patterns by team roles revealed that, in human-only teams, communication was longer and more frequent across all roles (Fig.~\ref{fig:behavior_fig2}g; frequency: ThrustPilot, $F(1, 20)=4.523, P = 0.0461$; YawPilot, $F(1, 20)=2.599, P = 0.1226$; PitchPilot, $F(1, 20)=3.414, P = 0.0795$; duration: ThrustPilot, $F(1, 20)=2.639, P = 0.1199$; YawPilot, $F(1, 20)=5.440, P = 0.0302$; PitchPilot, $F(1, 20)=3.918, P = 0.0617$). Notably, the ThrustPilot in human-only teams used command words more frequently, while other roles engaged in less frequent but longer communication. In contrast, within human-AI teams, the ThrustPilot (AI agent) communicated more frequently and for longer durations than human participants, indicating a shift in communication dynamics when AI is present.

As task difficulty increased, human-only teams exhibited more communication compared to human-AI teams (Fig.~\ref{fig:behavior_fig2}h; frequency: easy, $F(1, 20)=4.541, P = 0.0457$; intermediate, $F(1, 20)=2.993, P = 0.0990$; hard, $F(1, 20)=3.241, P = 0.0896$; duration: easy, $F(1, 20)=6.711, P = 0.0175$; intermediate, $F(1, 20)=6.914, P = 0.0161$; hard, $F(1, 20)=4.792, P = 0.0428$). Communication frequency and duration slightly increased with task difficulty in human-only teams but decreased in human-AI teams during more challenging tasks, suggesting that participants may focus more on individual performance when collaborating with AI.

The communication protocol significantly impacted communication patterns. More open communication protocols led to longer and more frequent interactions among team members. Human-only teams communicated more during command word and free-speaking protocols (Fig.~\ref{fig:behavior_fig2}j; frequency: command word, $F(1, 20)=3.253, P = 0.0864$; free speech, $F(1, 20)=5.042, P = 0.0362$; duration: command word, $F(1, 20)=5.825, P = 0.0255$; free speech, $F(1, 20)=7.167, P = 0.0145$). Both team types exhibited longer and more frequent communications as protocols became more open, indicating that communication protocols significantly influence patterns in both human-onlypupil dilation and constriction patterns  that reduced communication in human-AI teams may contribute to lower team performance. Effective communication is vital for team coordination and performance, and its reduction can hinder collaborative efforts. Research indicates that communication quality is often worse in human-AI teams compared to human-only teams, potentially leading to decreased performance \cite{kaelin2024developing, zhou2024impact}. In summary, the presence of an AI teammate alters communication dynamics, leading to less frequent and shorter interactions, which may contribute to reduced team performance. Understanding and addressing these communication challenges are essential for enhancing the effectiveness of human-AI collaborations.

\begin{figure}[htp]
    \centering
    \includegraphics[width=1\textwidth]{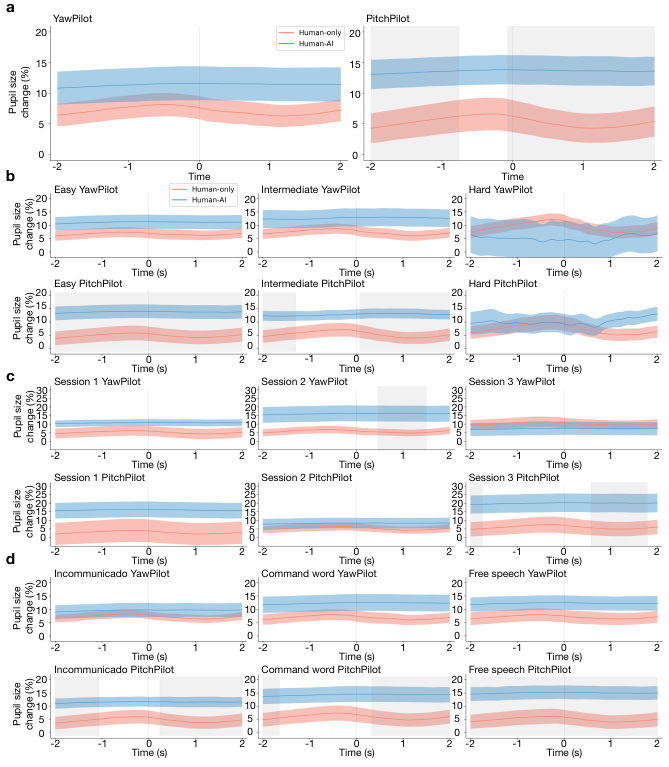}
    \caption{Pupil size changes of participants with different roles in human-only and human-AI teams (human-only: n=10 teams; human-AI: n=12 teams). \textbf{a}, Percent of pupil size change for participants with different roles across all task conditions. The color key for the team is used for b-d. The gray area indicates $p < 0.05$ by Welch’s t-test. \textbf{b-d}, Pupil size changes for different roles under various task conditions. \textbf{b}, Changes by difficulty level. \textbf{c}, Changes by experimental session. \textbf{d}, Changes by communication protocol.}
    \label{fig:pupil_fig3}
\end{figure}

\subsection{Pupil Dilation Indicated Higher Arousal in Human-AI Teams}
To assess differences in arousal when collaborating with AI agents versus humans, we analyzed percentage changes in pupil size from baseline during ring-passing events. Pupil size is a well-established marker of arousal levels \cite{nassar2012rational, joshi2016relationships}. In human-only teams, clear patterns of pupil dilation and constriction were observed (Fig.\ref{fig:pupil_fig3}a). Pupil size increased as participants approached the ring, constricted just before passing it, and dilated again afterward. However, in human-AI teams, participants exhibited higher overall pupil dilation across the entire epoch, reflecting sustained elevated arousal when working with an AI. Additionally, pupil dilation and constriction dynamics were less pronounced, suggesting a more stable yet heightened arousal state in human-AI teams. This effect was more pronounced for PitchPilots than YawPilots, highlighting role-specific variations in arousal dynamics  (see Section 7.1 in the Supplementary Information for more details).

\textbf{Pupil Dynamics Under Different Task Difficulty Levels}
We further investigated how task difficulty influenced pupil dynamics in human-only and human-AI teams, as previous studies suggest a positive correlation between pupil size and task difficulty \cite{bradley2008pupil, qin2022predictive}. As shown in Fig.~\ref{fig:pupil_fig3}b, under easy or intermediate tasks, participants in human-AI teams exhibited higher pupil dilation than those in human-only teams for both YawPilots (not significant, $P \geq 0.05$) and PitchPilots. This finding suggests heightened arousal in human-AI teams during less challenging tasks.

Interestingly, as task difficulty increased, pupil dilation consistently rose in human-only teams but did not show a similar pattern in human-AI teams. Instead, human-AI teams exhibited flattened pupil responses, potentially indicating overstimulation or disengagement as tasks became harder. These results suggest that task difficulty has a greater impact on arousal dynamics in human-only teams than in human-AI teams, possibly due to the human tendency to disengage or feel overwhelmed when collaborating with an AI agent under challenging conditions.

\textbf{Pupil Dynamics Under Different Experimental Sessions.}
Beyond arousal levels, pupil size reflects familiarity with task stimuli \cite{kafkas2012familiarity, franzen2022individual}. Over multiple sessions, participants in human-only teams demonstrated a slightly increasing pupil dilation, suggesting growing engagement and adaptation to the task. In contrast, this trend was less evident for participants collaborating with the AI agent (Fig.~\ref{fig:pupil_fig3}c). Moreover, in the first two sessions, both YawPilot and PitchPilot from human-AI teams exhibited higher pupil dilation compared to the two roles in human-only teams ($P \geq 0.05$). In the third session, only PitchPilot in the human-AI team showed higher pupil dilation.

Distinct pupil dynamics were observed across the sessions: human-only teams maintained a dilation-constriction-recovery pattern around each ring, whereas human-AI teams exhibited more uniform and flattened pupil responses. These differences highlight how the presence of an AI agent can disrupt typical patterns of arousal and adaptation in collaborative settings.

\textbf{Pupil Dynamics Under Different Communication Protocols.}
Pupil size changes from baseline were significantly larger for participants in human-AI teams compared to human-only teams, regardless of the communication condition (Fig.~\ref{fig:pupil_fig3}d). Consistent with prior observations, participants in human-only teams displayed the typical dilation-constriction-recovery pattern across all communication protocols, indicating stable arousal and engagement. In contrast, this pattern was absent in human-AI teams, suggesting disrupted arousal regulation when collaborating with an AI teammate.

The effect of communication protocols on pupil size was minimal in human-only teams. However, in human-AI teams, pupil size was notably larger when communication was allowed, potentially reflecting increased cognitive load or heightened arousal during open communication. These findings further emphasize the unique physiological responses elicited by AI teammates under varying communication conditions.

\subsection{Eye Blink Rates Were Higher in Human-AI Teams, Indicating Lower Engagement or Higher Cognitive Load}
\begin{figure}[htp]
    \centering
    \includegraphics[width=1\textwidth]{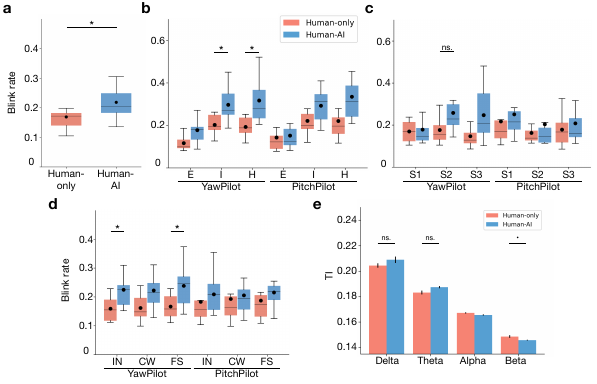}
    \caption{Blink rate and inter-brain synchrony between participants in human-only and human-AI teams. \textbf{a-d}, Blink rate of participants under different task conditions (human-only: n=10 teams; human-AI: n=12 teams). Black dots represent the means. \textbf{a}, Blink rate across all conditions. \textbf{b}, Blink rate of each role under different task difficulty levels. The color key for the team is used for c-d. \textbf{c}, Blink rate of each role within each experimental session. \textbf{d}, Blink rate of each role under different communication protocols. \textbf{e}, Inter-brain synchrony of different frequency bands (human-only: n=10 teams; human-AI: n=12 teams). In each team, inter-brain synchrony is measured using Total Interdependence (TI) between YawPilot and PitchPilot. One-way ANOVA with Bonferroni correction: ns. not significant,  $\cdot~P < 0.1, *~P < 0.05, **~P < 0.01, ***~P < 0.001.$}
    \label{fig:blink_eeg_fig4}
\end{figure}

As an indicator of attention and cognitive load \cite{chen2014using, maffei2019spontaneous, ranti2020blink}, eye blink rates significantly differed between participants in human-only and human-AI teams. Specifically, participants in human-AI teams exhibited higher blink rates than participants in human-only teams (Fig.\ref{fig:blink_eeg_fig4}a, $F(1, 20) = 4.988, P = 0.0371$). The higher blink rate suggests reduced engagement or increased cognitive load during collaboration with AI. Across various task difficulty levels, both YawPilots and PitchPilots in human-AI teams show higher blink rates (Fig.\ref{fig:blink_eeg_fig4}b, YawPilot: easy, $F(1, 20)=6.465, P = 0.0194$; intermediate, $F(1, 20)= 10.825, P = 0.0037$; hard, $F(1, 20)= 9.009, P = 0.0071$. PitchPilot: easy, $F(1, 20)= 0.077, P = 0.7844$; intermediate, $F(1, 20)= 4.513, P = 0.0463$; hard, $F(1, 20)= 5.432, P = 0.0304$.)

Notably, as shown in Fig.~\ref{fig:blink_eeg_fig4}c and d, both roles in human-AI teams had significantly higher blink rates than participants in human-only teams, regardless of the experimental session or communication protocol (YawPilot: session 1, $F(1, 20)=0.085, P = 0.7736$; session 2, $F(1, 20)= 4.811, P = 0.0403$; session 3, $F(1, 20)= 6.31, P = 0.0207$; incommunicado, $F(1, 20) = 7.730, P=0.0115$; command words, $F(1, 20) = 5.171,P = 0.0341$; free speech, $F(1, 20) = 7.966, P =0.0105$. PitchPilot: session 1, $F(1, 20)=0.287, P = 0.5982$; session 2, $F(1, 20)= 0.667, P = 0.4237$; session 3, $F(1, 20)= 0.526, P = 0.4768$; incommunicado, $F(1, 20) = 0.701, P =0.4123$; command words, $F(1, 20) = 0.097,P =0.7589$; free speech, $F(1, 20) = 0.852, P =0.3669$.) These findings suggest that collaboration with AI agents may reduce engagement or increase cognitive load among human participants, with this effect varying by participant role, session progression, and communication protocol.

Elevated blink rates may also correlate with lower team performance in human-AI teams. Increased blinking is often associated with reduced focus or increased cognitive load, which can detract from the attention required for precise coordination and task execution\cite{chen2014using, maffei2019spontaneous, ranti2020blink}. This heightened cognitive load in human-AI teams may lead participants to allocate more resources to manage their interaction with the AI, leaving fewer resources available for effective team collaboration. These results suggest that higher blink rates could serve as a physiological marker of reduced team efficiency in human-AI collaborations.

\subsection{Differences in EEG Synchrony as a Potential Marker of Team Performance}
We examined inter-brain synchrony through EEG analysis to evaluate neural coordination and arousal levels among team members during collaboration with either another human or a WOz AI agent \cite{dikker2017brain, chen2023inter, keynan2019electrical, vanhollebeke2023effects}. Inter-brain synchrony was quantified using Total Interdependence (TI), a measure of the mutual information shared between EEG signals across two participants, indicating the degree of neural coupling and coordination during team interactions. TI was analyzed across four frequency bands—delta (0.5–4 Hz), theta (4–8 Hz), alpha (8–13 Hz), and beta (13–30 Hz)—which reflect distinct aspects of neural activity related to arousal, attention, and cognitive engagement.

As shown in Fig.~\ref{fig:blink_eeg_fig4}e, participants in human-AI teams exhibited slightly higher TI in the delta and theta bands compared to human-only teams (delta: $F(1,20)=1.804, P =0.1943$; theta: $F(1,20)=5.757, P = 0.0263$). Delta and theta synchrony are often associated with cognitive engagement and shared attention during collaborative tasks \cite{balconi2015hemodynamic}, suggesting that human-AI teams may employ compensatory cognitive mechanisms to address the challenges of working with an AI teammate. In contrast, human-only teams exhibited slightly higher alpha and beta TI values, although these differences were not statistically significant. Alpha and beta synchrony are linked to focused attention, efficient information processing, and task-related cognitive efficiency \cite{popov2018time, chen2023inter}, which are crucial for high team performance.

The differences in EEG synchrony observed between human-only and human-AI teams suggest distinct underlying mechanisms driving team interactions. Higher delta and theta synchrony in human-AI teams may reflect an increased cognitive load or heightened coordination efforts to compensate for the AI’s limitations. Conversely, higher alpha and beta synchrony in human-only teams align with smoother collaboration and enhanced task efficiency, reinforcing the hypothesis that these bands are markers of effective team performance.

To further explore these relationships, we analyzed EEG synchrony under varying task conditions (see Supplementary Information 7.2 and 7.3). Additionally, we provide topographic maps of EEG activity for both team types, highlighting neural coupling across the frequency bands (Fig. S2). Correlation analyses between EEG synchrony and task performance revealed that higher alpha and beta synchrony were associated with better team performance in human-only teams, supporting the potential of EEG synchrony as a biomarker for team dynamics and effectiveness.


\subsection{Subjective Ratings of AI Were Initially Underrated but Improved Over Time}
After each experimental session, participants completed a questionnaire evaluating their team members, focusing on the ThrustPilot’s leadership, helpfulness, and trust in the AI agent (see \nameref{post_task_questionnaire} for details). This assessment allowed us to examine how collaborating with an AI agent influenced participants’ subjective perceptions over time.

\begin{figure}[htp]
    \centering
    \includegraphics[width=1\textwidth]{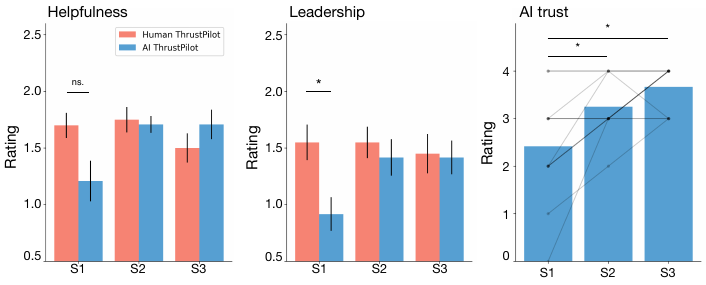}
    \caption{Subjective rating of ThrustPilot's helpfulness, leadership, and trust of the AI agent (human-only: n=10 teams; human-AI: n=12 teams). One-way ANOVA with Bonferroni correction and repeated measures ANOVA for AI trust, ns., not significant; $*~P < 0.05$.}
    \label{fig:questionnaire_fig5}
\end{figure}
In the first experimental session, ratings for the AI agent were significantly lower than those for human participants in the same role (Helpfulness: $F(1,20)=4.948, P=0.0378$; Leadership: $F(1,20)=8.510, P=0.0085$; Fig.~\ref{fig:questionnaire_fig5}). These initial findings suggest that participants undervalued the AI agent’s contributions early in the collaboration.

Over time, subjective ratings for the AI agent improved significantly. By Session 3, the differences in ratings for the ThrustPilot controlled by a human or the WOz AI became negligible, indicating growing acceptance of the AI agent’s role. Trust in the AI agent also increased significantly across sessions ($F(2,22)=10.405, P=0.0007$), with notable improvements from Session 1 to Session 2 ($P=0.0104$) and from Session 2 to Session 3 ($P=0.0045$). This progression suggests that repeated interactions foster greater trust in AI collaborators, reflecting a learning curve in human-AI partnerships.

These findings highlight the potential of extended interaction to mitigate initial biases against AI agents. As participants spent more time collaborating with the AI, they appeared to better understand and appreciate its contributions to the team. This growing trust and improved evaluation underscore the importance of designing AI systems that support sustained and meaningful interactions to enhance collaboration quality over time.

\section{Discussion}\label{discussion}
In this study, we investigated the impact of integrating an AI agent into human teams on team performance, dynamics, and individual physiological and behavioral responses. Strikingly, our findings reveal that teams collaborating with an AI agent performed worse than human-only teams, with the performance gap becoming most pronounced during tasks requiring high levels of coordination. This counterintuitive result challenges assumptions that AI integration inherently enhances team outcomes. By analyzing behavioral patterns, communication dynamics, and physiological responses, we identified key disruptions introduced by the AI agent, underscoring the complexities and unintended consequences of incorporating AI into collaborative settings.


Our findings show that overall team performance was lower in human-AI teams compared to human-only teams, a pattern that persisted across different task difficulty levels and was not influenced by communication protocols or the number of experimental sessions. While studies suggest AI can enhance decision-making and team performance in certain contexts \cite{flathmann2023examining, passalacqua2024practice}, our results demonstrate that in tasks requiring rapid sensorimotor decision-making and intensive collaboration, the presence of an AI agent—when identified as such by participants—detrimentally affects performance. Interestingly, participants’ subjective perceptions of the AI teammate improved over time, as measured by post-task surveys, but this trust did not translate into better team performance. This disconnect may stem from the AI’s inability to develop a shared mental model with human teammates, which is critical for effective collaboration \cite{fiore2010toward}. Fig.~\ref{fig:experiment_fig1} and Fig.~\ref{fig:questionnaire_fig5} show participants’ increasing trust in the stable performance gap, emphasizing that trust alone is insufficient to overcome coordination inefficiencies caused by the AI’s limited adaptability to human behavior. Future research should investigate how trust, shared mental models, and AI adaptability interact to design AI agents that enhance both team dynamics and performance in challenging tasks.

Verbal communication plays a pivotal role in team dynamics, enabling the exchange of information, strategic discussions, and emotional support. Different communication protocols, such as command word communication and free speech, offer varying levels of efficiency and openness. However, introducing an AI agent can significantly alter these communication patterns \cite{johnson2020training}. Previous research has indicated that communication among human team members often decreases in the presence of AI agents \cite{johnson2020training, shaikh2023ai}. Although participants may be open to AI collaboration, ensuring effective and neutral language communication remains a challenge in human-AI teams \cite{zhang2021ideal, musick2021happens}. In our experiment, participants communicated less frequently and more briefly in human-AI teams than in human-only teams, suggesting a reluctance to engage with AI team members or other humans when AI is involved. Developing AI agents that can communicate more naturally and fluidly with human counterparts is critical to fostering effective collaboration \cite{demir2015synthetic}. Our findings emphasize the need to improve AI communication capabilities to ensure smoother interactions in human-AI teams.

Behavioral and physiological measures, such as the number of remote controller actions, pupil size, and blink rate, all suggest that participants experienced higher arousal when collaborating with an AI agent than when working solely with human team members. According to the Yerkes–Dodson law, performance is related to arousal levels \cite{yerkes1908relation}. Previous studies have found that humans experience greater cognitive load when working with AI agents, leading to reduced team efficiency \cite{mcneese2018teaming, demir2018team, flathmann2023examining}. This points to the need for AI systems that can help regulate human arousal levels. By modifying their behavior or through communication, AI agents could help maintain an optimal level of arousal to maximize team performance. This approach offers a new direction for designing AI agents in human-AI collaborative settings.

Previous research has demonstrated a positive correlation between inter-brain synchrony and team performance \cite{barraza2020brain, wikstrom2022inter}. Our study supports this, showing that including the WOz AI agent, Alice, led to reduced alpha and beta synchronization, which was associated with lower team performance. This suggests that brain synchrony could serve as a potential biomarker for assessing team dynamics and performance \cite{gordon2020physiological, li2021dynamic}. Our goal was to explore how the presence of an AI agent affects neural synchrony within human teams and, consequently, overall performance. The observed decrease in alpha and beta band synchrony highlights the intricate relationship between neural synchrony and collaboration. These findings underscore the importance of considering neural markers in the design of AI systems intended for human-AI collaboration. Future research should investigate how different AI agent characteristics influence brain synchrony and develop strategies to enhance synchrony and improve team performance in human-AI settings.

In conclusion, our study reveals that the inclusion of AI in human teams poses significant challenges to team dynamics, communication, physiological responses, and overall performance. As AI becomes more prevalent in collaborative environments, it is crucial to continue refining how AI agents interact with humans to support efficient and productive teamwork.

\section{Methods}\label{methods}

\subsection{Participants}
Ninety-five participants were recruited to participate in three sessions of the experiment in either dyad or triad teams (13 dyad teams and 23 triad teams). One dyad team and five triad teams were excluded due to at least one participant being unable to complete all three experimental sessions. An additional eight triad teams were excluded due to technical issues with either desktop systems or EEG devices. Ten triad teams and twelve dyad teams were included in the final analysis (29 male, 24 female, and 1 non-binary; mean age $=23.15$ years, age standard deviation $=2.73$ years). All participants had normal or corrected-to-normal vision and provided informed consent before each experiment. The protocol was approved by the Columbia University Institutional Review Board.

\subsection{Procedure}
Upon arrival for the first session, participants ($N=2$ for human-AI teams, $N=3$ for human-only teams) watched an instructional video together. Following this, EEG devices were set up, and participants were escorted to separate EEG recording chambers with soundproofing to block external noise and minimize electrical interference. Once settled, participants were equipped with head-mounted displays (HMDs) and remote controllers, followed by an eye calibration procedure. Participants then completed five pilot trials, where no data were collected, to allow for hardware testing and familiarization.

Data collection began immediately after the pilot trials, randomly assigning roles to each participant. Every five trials, participants received instructions through their headphones regarding the communication protocol for the next set of trials. Each trial ended if the spacecraft collided with a ring, missed passing through it, or the team ran out of time. Upon completion of each trial, participants were shown either ``Success'' or ``Failure'' on their HMDs.

All teams completed three experimental sessions, with at least 24 hours between each. None of the participants had prior experience with the task before the first session. After each session, participants completed a post-task questionnaire (details provided in \nameref{post_task_questionnaire}).

\subsection{Apollo Distributed Control Task}
The Apollo Distributed Control Task (ADCT) required three participants to collaborate on navigating a spacecraft within a limited amount of time. Each participant controlled a different degree of movement and had a unique first-person view of the environment. Specifically, the YawPilot controlled left-right movements, the PitchPilot controlled up-down movements, and the ThrustPilot controlled the speed of the spacecraft.

Task difficulty was divided into three levels, determined by the radius of the rings the team needed to navigate through. Smaller ring radii made the task more difficult. Each trial consisted of 15 rings with varying radii. Easy rings had a radius 2.5 times the spacecraft's radius, intermediate rings had a radius 2 times the spacecraft's radius, and hard rings had a radius 1.5 times the spacecraft's radius. Difficulty increased every five rings, progressing from easy to hard.

ADCT allowed for three communication protocols: 1) Incommunicado, where no communication between participants was permitted; 2) Command Word Communication, where participants could communicate using specific command words (e.g., ``up,'' ``down,'' ``left,'' ``right,'' ``faster,'' ``slower''); and 3) Free Speaking, where participants could communicate freely. In communication-allowed conditions, multiple participants could speak simultaneously.

\subsection{Alice the Wizard of Oz (WOz) AI Agent}
The WOz AI agent, named ``Alice,'' was introduced to participants via an instructional video, where it was described as a state-of-the-art AI agent capable of real-time communication and collaboration in the ADCT task. Alice's voice was modified using a vocal transformer (Roland VT-4) to prevent participants from detecting that Alice was human-controlled.

To ensure consistency across trials, the same experimenter played the role of Alice for all teams. Alice controlled the spacecraft using a 32-inch monitor with a 1920x1080 resolution and 75Hz refresh rate instead of a head-mounted display. Only Alice's behavioral data (speech and keyboard actions) were recorded during the experiment.

\subsection{Virtual Environment and Task Setup}
The virtual environment for ADCT was developed using Unreal Engine (version 4.24.3). It included a spacecraft, transparent rings, Earth, and a time bar. The spacecraft had three windows corresponding to the views for YawPilot, PitchPilot, and ThrustPilot, with each role seeing a unique perspective of the environment.

Participants had to guide the spacecraft through transparent red rings spaced uniformly throughout the environment, with the rings serving as the primary task stimuli. Earth was placed at the end of each trial to signal completion. A time bar, displayed at the bottom of each window, tracked the time remaining in each trial.

A random 5Hz turbulence was introduced to the spacecraft’s yaw and pitch movement to increase task difficulty and engagement. This turbulence required participants to adjust the spacecraft’s trajectory continually, fostering constant coordination and interaction.

\subsection{Remote Controller Actions}
Participants used VIVE Pro Eye controllers to adjust the spacecraft’s movement. The YawPilot controlled left-right movement, while the PitchPilot and ThrustPilot used the up-down controls on the trackpad. The number of remote controller actions (i.e., trackpad presses) was logged, with each press, even if held, counted as one action.

\subsection{Head-Mounted Display and Eye Tracking}
VIVE Pro Eye HMDs were used, with a combined resolution of 2880x1600, a 90Hz refresh rate, and a 110-degree field of view. The built-in eye tracker recorded pupil size and eye openness at 120Hz.

Pupil size data were preprocessed by removing artifacts, downsampled to 60Hz, and filtered using a fourth-order low-pass Butterworth filter with a 4Hz cutoff. Data were then segmented into 4-second epochs centered on each ring-passing event. The blink rate was calculated from eye openness, with eye openness below 30\% classified as a blink.

\subsection{Speech Recording and Analysis}
Speech data were recorded using USB microphones (TONOR TC-777) and VIVE Pro Eye’s built-in headphones. Alice’s communication was played through a JBL GO 2 speaker. All speech data were synchronized using Lab Streaming Layer (LSL) and preprocessed to remove background noise using the \textit{noisereduce} Python library. Voice activity detection was applied to detect speech events, which were then downsampled and segmented into epochs around ring-passing events.

Communication frequency was calculated as the number of speech events per epoch, while communication duration was defined as the total time spent communicating within each epoch.

\subsection{EEG Data Acquisition and Preprocessing}
EEG data were recorded using the Advanced Brain Monitoring B-Alert X24 system, which featured 20 channels positioned according to the international 10-20 system. Data were sampled at 256Hz and preprocessed using the \textit{MNE} Python package \cite{GramfortEtAl2013a}. Bad channels were removed, and Independent Component Analysis (ICA) was performed to remove artifacts from eye blinks and muscle activity. The cleaned EEG data were epoched into 4-second intervals around ring events for further analysis.

\subsection{EEG Synchrony}
To assess inter-brain synchrony between YawPilot and PitchPilot, Total Interdependence (TI) was computed, a measure of coherence between pairs of EEG channels from different participants \cite{wen2012exploring, dikker2017brain, bevilacqua2019brain}. Spectral coherence was calculated using the Welch method for different frequency bands: delta (0.5-4Hz), theta (4-8Hz), alpha (8-13Hz), and beta (13-30Hz). TI was computed for each epoch and averaged across all channel pairs to assess global brain-to-brain synchrony.

\subsection{Post-task Questionnaire}
\label{post_task_questionnaire}
After each experimental session, participants completed a post-task questionnaire. Two key questions focused on participants' subjective evaluations of their teammates:
\begin{enumerate}
    \item How helpful was each of your teammates in reaching the final solution?
    \item How much did each of your teammates act as a leader in today’s experiment?
\end{enumerate}
Each teammate was rated on a three-point scale: ``Not at all," ``A little," or ``Very well". 

\section{Acknowledgments}
This work was supported by funding from the Army Research Laboratory’s Human-Guide Intelligent Systems (HGIS) Program (W911NF-23-2-0067) the Army Research Laboratory’s STRONG Program (W911NF-19-2-0139, W911NF-19-2-0135, W911NF-21-2-0125)  the National Science Foundation (IIS-1816363, OIA-1934968) the Air Force Office of Scientific Research (FA9550-22-1-0337) and a Vannevar Bush Faculty Fellowship from the US Department of Defense (N00014-20-1-2027).

\section{Data Availability}
All de-identified data and code used for generating the results presented in this paper are publicly available at \href{https://github.com/YinuoQ/human_AI_behavioral_and_physiological_data}{https://github.com/YinuoQ/human\_AI\_behavioral\_and\_physiological\_data}. This repository includes documentation detailing the code, instructions for replicating the analysis, and link of the dataset. 
\bibliography{sn-bibliography}

\section{Supplementary Information}

\subsection{Team-based pupil size analysis}
\begin{figure}[htp!]
    \centering
    \includegraphics[width=0.9\textwidth]{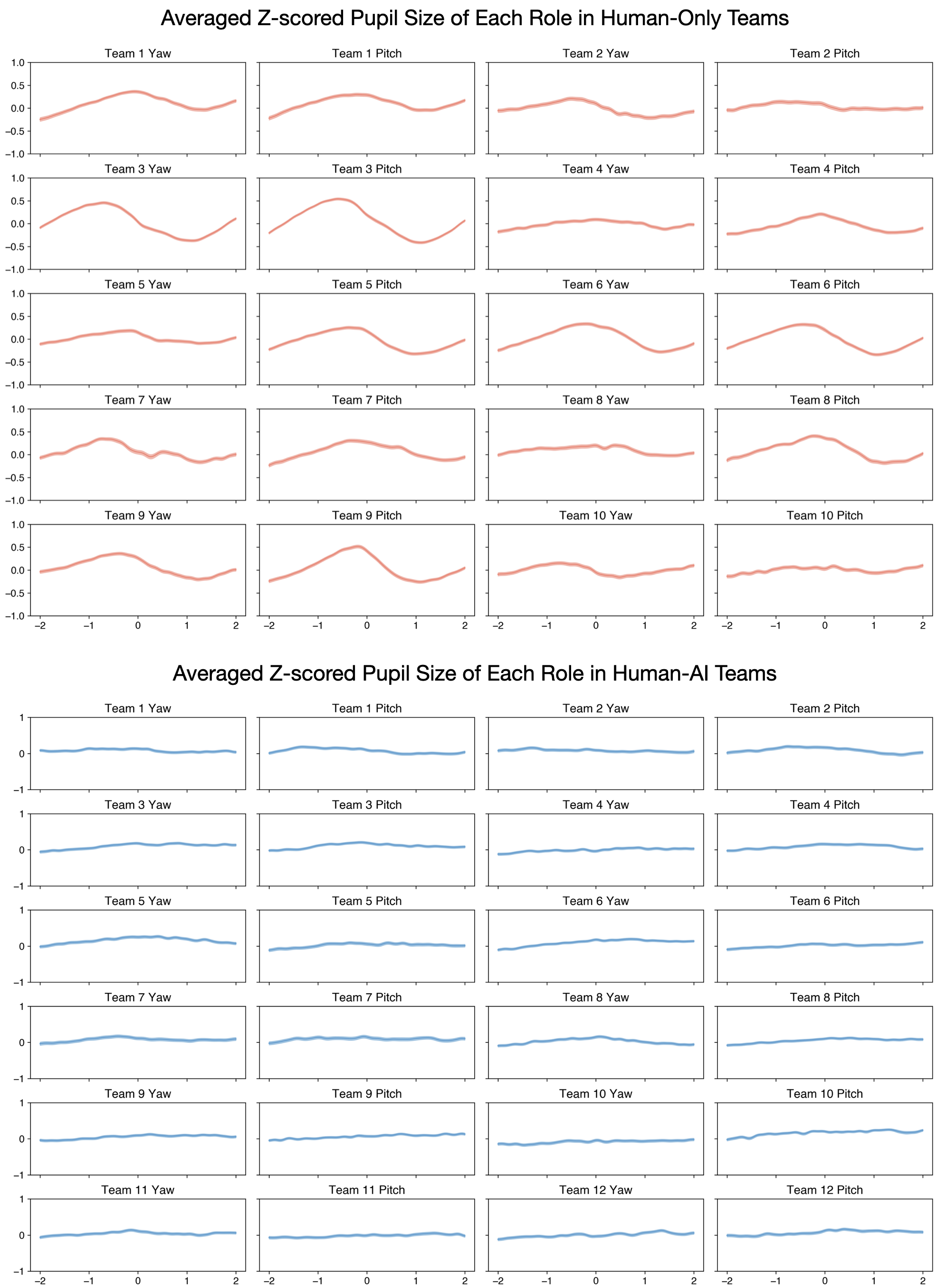}
    \caption{Averaged z-scored pupil size across teams for human-only and human-AI teams. Each subplot represents the averaged z-scored pupil size of individuals in specific roles. Each epoch is centered around passing a ring, with 2 seconds before and after the event. Pupil size changes are z-scored based on trials, and the shaded areas represent the Standard Error of the Mean (S.E.M.). The top panel shows human-only teams, where larger fluctuations in pupil size are observed, while the bottom panel illustrates human-AI teams, which exhibit flatter pupil size trends.}
    \label{fig:supp1_pupil}
\end{figure}

Fig.~\ref{fig:supp1_pupil} presents the z-scored pupil size dynamics across each team. As an extension of Fig.~\ref{fig:pupil_fig3}, which demonstrates overall pupil percent changes, Fig.~\ref{fig:supp1_pupil} employs z-scored pupil size to emphasize within-team variations, offering a deeper understanding of how individual participants' pupil size fluctuate relative to their own baseline. Specifically, z-scored data helps normalize individual differences. Our results provide a clear temporal view of how participants' arousal levels fluctuate around the ring-passing event.

In human-only teams, we observe more pronounced fluctuations in pupil size, consistent across multiple teams and roles. This variability indicates high arousal changes around the ring. The pupil dynamics suggest that participants may feel nervous as they approach the ring, and their arousal levels recover after passing or believing they can pass it. In contrast, participants in human-AI teams exhibit more stable pupil responses. Combined with the results from Fig.~\ref{fig:pupil_fig3}, which shows that the overall percentage increase in pupil size from baseline is significantly higher when collaborating with an AI agent, the flatter pupil dynamics suggest higher and more stable arousal levels during AI collaboration. This group-based z-scored data highlights how different team compositions influence physiological responses, indicating that working with an AI agent may create continuous and higher pressure for human team members. 

By analyzing the z-scored pupil data, subtle differences between individuals become more apparent compared to simply averaging pupil percent changes across teams. The z-scored approach highlights how individuals' pupil sizes fluctuate in response to passing a ring. The smaller fluctuations in z-scored pupil size for participants in different roles within human-AI teams suggest that the arousal level changes are less pronounced compared to participants in human-only teams. This reduced variability holds across multiple teams and participants, indicating more uniform and stable arousal dynamics in human-AI collaboration.

\newpage
\subsection{Topographies of Participants in Different Teams}
\begin{figure}[htp!]
    \centering
    \includegraphics[width=1\textwidth]{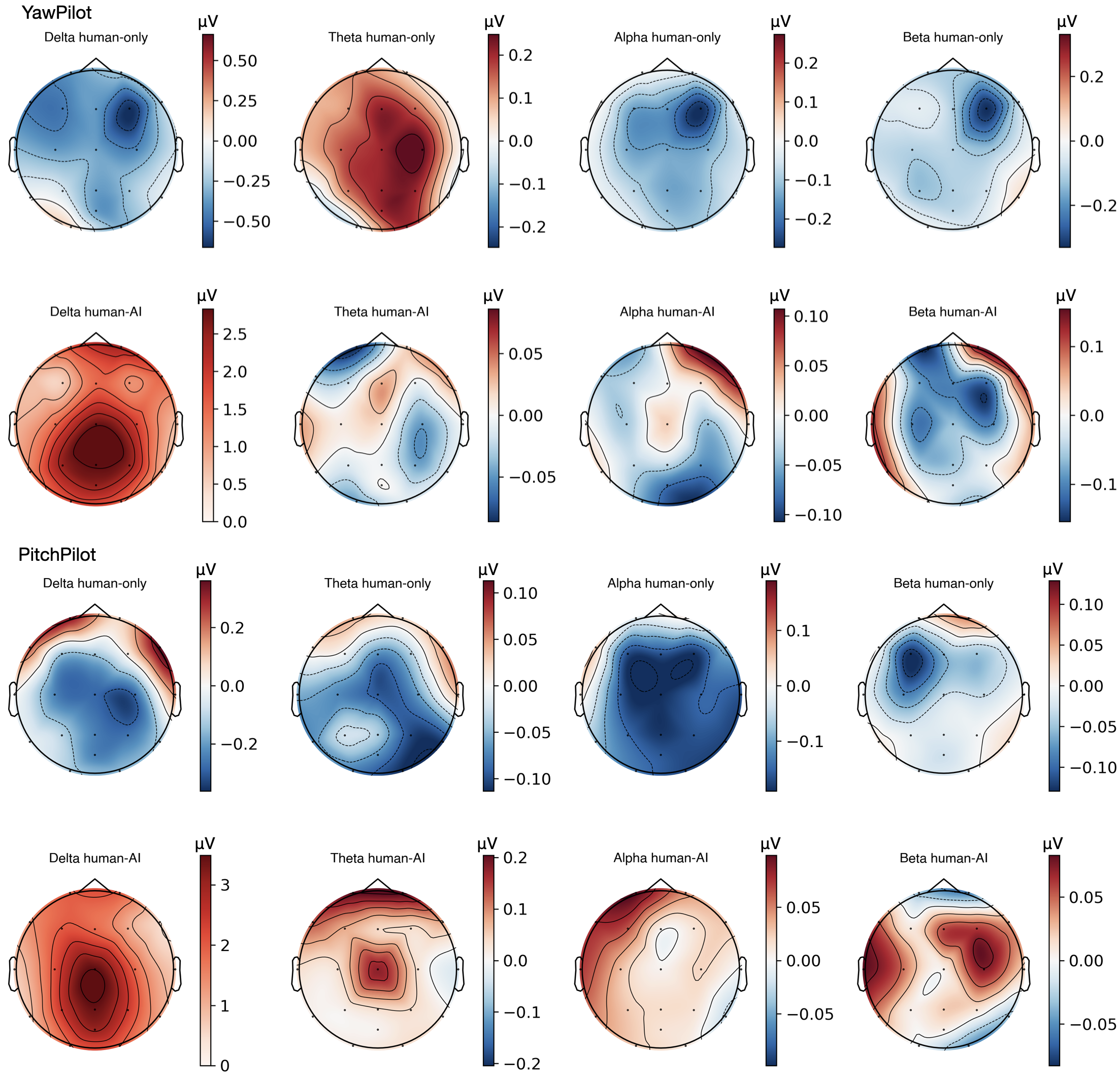}
    \caption{Topographic maps of the delta, theta, alpha, and beta bands when the team passes a ring for YawPilot and PitchPilot in human-only and human-AI teams.}
    \label{fig:supp3_eeg}
\end{figure}
The topographic maps present distinct patterns of neural activity across the delta, theta, alpha, and beta frequency bands for participants in different team configurations: human-only versus human-AI teams. As shown in Fig.~\ref{fig:supp3_eeg}, in the delta band, higher amplitude exists in the human-AI teams compared to human-only teams for both roles, especially in the central regions. Delta power is often associated with attention and cognitive control, with stronger delta activation in human-AI teams potentially indicating increased demands for monitoring and adapting to the AI agent. This suggests a higher arousal state of participants when collaborating with AI. Theta band activity is associated with cognitive control and arousal level. We observed a higher amplitude of theta activity of YawPilot in the human-only teams and PitchPilot in human-AI teams in the central area. The higher amplitude may indicate higher engagement and focused attention. 

In the alpha and beta bands, distinct patterns emerge that differentiate participants in human-only teams from those in human-AI teams. For the alpha band, human-only teams exhibit reduced amplitude across parietal regions, which is commonly associated with higher task engagement and attentional focus. In contrast, human-AI teams show elevated alpha amplitude. These results suggest that participants are more immersed in the task when interacting with human teammates than WOz AI. Similarly, in the beta band, human-only teams display lower activity in parietal regions, indicating sustained attention and steady task engagement. 

\newpage
\subsection{Inter-Brain Synchrony Under Different Task Conditions}
\begin{figure}[htp!]
    \centering
    \includegraphics[width=1\textwidth]{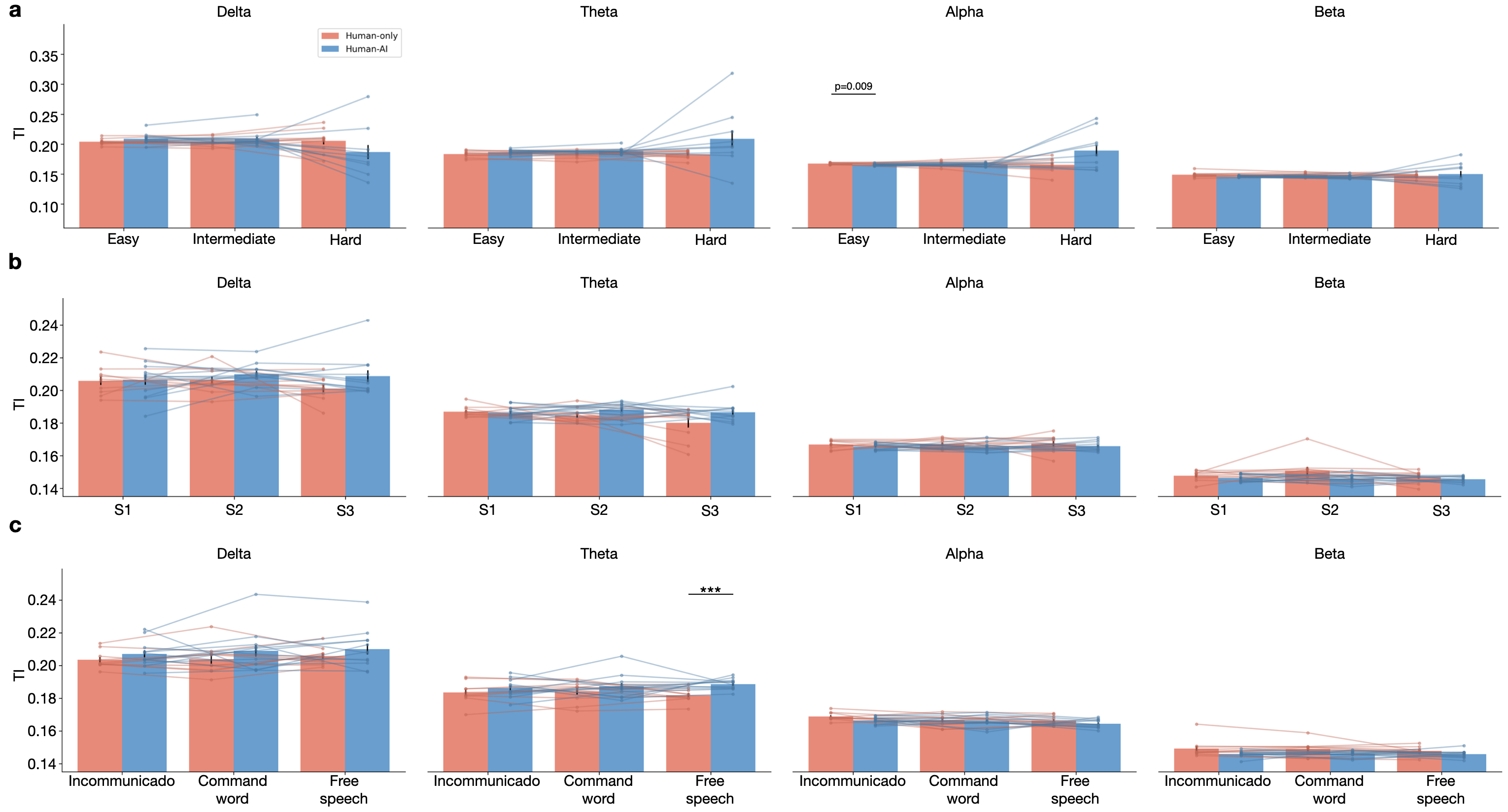}
    \caption{Inter-brain synchrony of different frequency bands under different task conditions (human-only: n=10 teams; human-AI: n=12 teams). In each team, inter-brain synchrony is measured using Total Interdependence (TI) between YawPilot and PitchPilot. \textbf{a}, Brain-synchrony under different task difficulty levels and frequency bands. Color key for team is used for b and c. (Delta: easy, $P = 0.1280$; intermediate, $P = 0.3718$; hard, $P = 0.2393, n=9$. Theta: easy, $P = 0.0902$; intermediate, $P = 0.0228$; hard, $P = 0.1257, n=9$. Alpha: easy, $P = 0.0095$; intermediate, $P = 0.6612$; hard, $P = 0.0443, n=9$. Beta: easy, $P = 0.0432$; intermediate, $P = 0.0331$; hard, $P = 0.6845, n=9$). \textbf{b}, Brain-synchrony within each experimental session and frequency bands (Delta: session 1, $P = 0.8834$; session 2, $P = 0.2606$; session 3, $P = 0.1218$. Theta: session 1, $P = 0.6689$; session 2, $P = 0.0781$; session 3, $P = 0.0882$. Alpha: session 1, $P = 0.2111$; session 2, $P = 0.1205$; session 3, $P = 0.4110$. Beta: session 1, $P = 0.1994$; session 2, $P = 0.0502$; session 3, $P = 0.2526$). \textbf{c}, Brain-synchrony under different communication protocols and frequency bands (Delta: incommunicado, $P = 0.2297$; command words, $P = 0.2956$; free speech, $P = 0.3150$. Theta: incommunicado, $P = 0.3196$; command words, $P = 0.3145$; free speech, $P = 0.0004$. Alpha: incommunicado, $P = 0.0212$; command words, $P = 0.7138$; free speech, $P = 0.1212$. Beta: incommunicado, $P = 0.0673$; command words, $P = 0.0431$; free speech, $P = 0.0557$). One-way ANOVA with Bonferroni correction $***P < 0.001.$}
    \label{fig:supp4_TI}
\end{figure}
\label{TI under different conditions}
We examined how task difficulty influenced inter-brain synchrony across human-only and human-AI teams (Fig.~\ref{fig:supp4_TI} a). Delta and theta band TI were significantly lower at easier task levels in human-only teams, suggesting that human-AI teams experience stronger neural coupling even in simpler tasks \cite{valencia2020binds}. In intermediate and hard tasks, human-AI teams continued to show higher TI in the theta band, while human-only teams exhibited higher alpha and beta TI. These findings suggest that human-AI collaborations may promote greater engagement in more cognitively demanding tasks, yet this heightened synchrony does not always translate to improved performance.

TI was also analyzed across three experimental sessions to assess how familiarity with the task and teammates influenced inter-brain synchrony (Fig.~\ref{fig:supp4_TI} b). In the first session, there were no significant differences between human-only and human-AI teams across all frequency bands. However, human-only teams exhibited higher alpha and beta TI by the second session, while human-AI teams showed increased delta and theta synchrony. These differences became more pronounced in the third session, suggesting that human teams develop stronger synchrony over time, while human-AI teams maintain higher levels of synchrony in lower-frequency bands, potentially reflecting a different form of cognitive or emotional engagement.

Finally, we analyzed TI across different communication protocols (Fig.~\ref{fig:supp4_TI} c). In free speech trials, human-AI teams showed significantly higher theta TI than human-only teams, indicating that open communication enhances neural synchrony in these teams. In contrast, human-only teams exhibited higher alpha and beta TI during incommunicado trials. This suggests that non-verbal coordination is more significant in human-human teams when communication is restricted. These results underscore the importance of communication strategies in shaping inter-brain synchrony in human-AI collaboration.
\end{document}